\documentstyle[preprint,aps,epsfig]{revtex}

\begin{document}
\draft
\preprint{}
\def\qslash{\hbox{q\kern-.5em\lower.1ex\hbox{/}}}


\begin{flushright} 
USM-TH-75
\end{flushright}

\bigskip\bigskip
{\centerline{ \bf\huge QCD condensate contributions
to the effective}
{\centerline{\bf\huge  quark potential in a covariant gauge }}

\vspace{22pt}

\centerline{
 {Iv\'an Schmidt$^1$ and 
Jian-Jun Yang $^{1,2}$}}

\vspace{8pt}
{\centerline{\it
$^1$ Departamento de F\'{\i}sica, Universidad T\'ecnica 
Federico Santa Maria,} 
{\centerline {\it Casilla 110-V, Valpara\'{\i}so,  Chile}}

{\centerline{\it 
$^2$ Department of Physics, Nanjing Normal  University, 
Nanjing 210097, P. R. China}

\vspace{10pt}
\begin{center} 
{\large \bf Abstract}
\end{center}

We discuss QCD condensate
contributions to the gluon propagator
both in the fixed-point gauge and
in covariant gauges for the external
QCD vacuum gluon fields with the conclusion that
a covariant gauge is essential to obtain a gauge invariant
QCD vacuum energy density difference and to retain the
unitarity of the quark scattering amplitude. The
gauge-invariant QCD condensate contributions to
the effective one-gluon exchange potential are evaluated
by using the effective gluon propagator  which produces
a gauge-independent quark scattering amplitude.

\vspace{1cm}

\noindent 
PACS number(s): 12.39.Pn, 12.38.Aw, 12.38.Lg

\newpage

{\centerline {\bf I.
INTRODUCTION}}

\medskip

The  quark potential model based on the one-gluon exchange
approximation can  correctly reproduce the baryon spectrum and
the static properties of
hadrons \cite{Spect}, especially the
positive parity states; including  the $q \bar{q}$ 
creation and annihilation terms in the one-gluon-exchange
approximation, it
can also  give a description of meson-nucleon interactions \cite{Muller}. 
Nevertheless, it is well understood that  the one-gluon exchange 
can only generate the short-range part of the baryon-baryon 
interaction since the one-gluon exchange potential is the 
nonrelativistic reduction of the  operator derived in the 
perturbative QCD scheme. It is clear that, to reflect medium- and
long-range QCD, some nonperturbative effects induced by the 
complicated structure of the QCD vacuum should be taken into 
account. That the QCD condensates could produce the 
nonperturbative correction to the traditional   potential model 
was first suggested by Shen et al. \cite{SPN}, with inspiration 
of QCD sum rules \cite {SVZ}. Several papers have appeared
afterwards studying  the QCD condensate contributions
to the quark potential\cite{Liu,Ding}. Possible
effects of QCD condensates in heavy quarkonium spectra were also
investigated by these authors. Their  results indicate that the 
nonperturbative effects induced by QCD vacuum condensates play a 
significant  role in the corrections to the $1/q^2$ behavior.

Unfortunately, the authors of the cited papers obtained the quark
potential from a gauge dependent quark-quark scattering
amplitude.
Nonperturbative effects  were
phenomenologically considered
in Refs. \cite{SPN,Liu,Ding} by
employing the vacuum condensate to modify the free gluon propagator
in the fixed-point gauge, i.e., the non-local
two-quark and  two-gluon vacuum expectation
values (VEVs) were calculated in  the fixed-point
gauge \cite{Schwinger}. Although this gauge is extremely simple
for many lowest order expansions\cite{Novikov}, it
violates translational invariance, and could in principle
conflict with the covariant gauge used to formulate QCD.
In particular, an explicit $\xi$-dependence
(here and henceforth, we
have specified the perturbative gauge dependence as
$\xi$-dependence) in
the transverse portion of the nonperturbative
gluon propagator results in a $\xi$-dependent
quark-quark scattering amplitude.
The effective quark potential 
obtained from a $\xi$-dependent amplitude
is of course $\xi$-dependent too.

The aim of this paper is twofold: (i) to
discuss QCD condensate
contributions to the gluon propagator
and, (ii) to evaluate the non-perturbative
corrections to the one-gluon exchange quark potential in covariant  
gauges.  This is then a natural continuation of the work
done in Ref. \cite{JJYang}. We use here the
nonperturbative gluon propagator,
allowing for the presence of the ghost condensate which
appears in covariant gauges.

In Sec.~II, we  discuss
the nonperturbative gluon propagator both 
in the fixed-point gauge and in  covariant gauges.
In Sec.~III, we present our main results of
the  gauge-invariant quark potential with the
nonperturbative corrections of QCD vacuum condensates.
The brief summary in Sec.~IV contains some possible
applications of the obtained quark potential.

\vspace{0.5cm}

{\centerline {\bf II. NONPERTURBATIVE GLUON  PROPAGATORS}}
{\centerline {\bf IN DIFFERENT GAUGES }}

\medskip

In this section, we discuss the
nonperturbative gluon propagator in the fixed-point
gauge and in covariant gauges for the external QCD
vacuum gluon fields (nonperturbative ones).
Recently, the operator product expansion (OPE) of
the gluon propagator in QCD has been extensively
studied
\cite{Lavelle1,Lavelle2,Lavelle3,Lavelle4,Lavelle5,Bagan}.

In QCD sum rules for gauge invariant currents,
the background field method is used, where
the fixed-point gauge is generally employed for
nonperturbative gluon fields,
i.e.,

\begin{equation}
x_\mu B^{\mu}_a(x)=0.
\end{equation}
But, for perturbative gluon fields, a covariant gauge
is usually adopted , i.e.

\begin{equation}
iD_{\mu\nu}^{ab}(q)=i \delta_{ab} \left [-\frac{g_{\mu\nu}}{q^2}+(1-\xi)\frac{q_{\mu}q_{\nu}}
{(q^2)^2}\right ].
\end{equation}
In the fixed-point gauge, the non-local two-gluon  VEV  
can be written as \cite{SVZ,Elias}
\begin{eqnarray}
\langle 0|B^a_\mu(x)B^b_\nu(y)|0 \rangle &=& \frac{1}{4}x^\rho y^\sigma
\langle 0|G^a_{\rho \mu}G^b_{\sigma \nu} |0 \rangle
  + \cdots \cdots  \nonumber \\
  &=& \frac{\delta_{ab}}{48(N_c^2-1)} x^\rho y ^\sigma (g_{\rho \sigma}
  g_{\mu \nu}-g_{\rho \nu} g_{\sigma \mu})
  \langle 0 | G^2 | 0 \rangle +\cdots, \label{BB1}
\end{eqnarray}
where
\begin{equation}
  \langle 0 |G^2 | 0 \rangle=\langle 0|G^a_{\rho \mu}G_a^{\rho \mu} |0 \rangle.
\end{equation}
Obviously, the expansion (\ref{BB1}) violates translational
invariance since the right hand side(RHS) of (\ref{BB1}) is
a function of $xy$ instead of $(x-y)$.

To  obtain the correct nonperturbative gluon propagator, it is essential to
obtain the expansion of
$\langle 0|B^a_\mu(x)B^b_\nu(y)|0 \rangle$ with
translational invariance.
The basic requirements for translational
invariance were studied before \cite{YZPC5}.
According to  these requirements
and the Lorentz gauge condition 
\begin{equation}
\partial ^\mu B^a_\mu (x)=0, \label {LC}
\end{equation}
the non-local two-gluon VEV can be expressed as

\begin{eqnarray}
& &\langle 0 | B^a_\mu (x) B^b_\nu (y) | 0 \rangle =
\langle 0 | B^a_\mu (0) B^b_\nu (0) | 0 \rangle \nonumber \\
& &- \frac{\delta_{ab}}{2(N_c^2-1)} (x-y)^{\rho} (x-y)^{\sigma} 
\langle 0 | \partial _\rho B^d_\mu (0)\partial _\sigma  
B^d_\nu (0) | 0 \rangle +\cdots \cdots, \label {BB6}
\end{eqnarray}
where
\begin{eqnarray}
\langle 0 | B^a_\mu (0) B^b_\nu (0) | 0 \rangle &=&
\frac{g_{\mu \nu}}{4} \frac{\delta_{ab}}{(N_c^2-1)} 
\langle 0 | B^d_\rho (0) B^{\rho}_d (0) | 0 \rangle \nonumber \\
&=& 
\frac{g_{\mu \nu}}{4} \frac{\delta_{ab}}{(N_c^2-1)} 
\langle 0 | B^2 | 0 \rangle \label {BB7}
\end{eqnarray}
and 
\begin{equation}
\frac{1}{2} \langle 0 | \partial_\rho B^a_\mu (0) \partial_\sigma  
B^a_\nu (0) | 0 \rangle = \left [S g_{\mu \nu} g_{\rho \sigma }
+\frac{R}{2}( g_{\rho \nu }g_{\sigma \mu}
+ g_{\rho \mu }g_{\nu \sigma})\right ]. \label{BB8}
\end{equation}
Contracting  Eq. (\ref {BB8}) with $g^{\rho \sigma} g^{\mu \nu}$,
$g^{\rho \mu} g^{\sigma \nu}$ and
$g^{\rho \nu}g^{\sigma \mu}$
leads to
\begin{equation}
\frac{1}{2} \langle 0 | \partial^\sigma B_a^\nu (0) \partial_\sigma  
B^a_\nu (0) | 0 \rangle = 16S+4R,        \label {CT1}
\end{equation}
 
\begin{equation}
\frac{1}{2} \langle 0 | \partial^\mu B^a_\mu (0) \partial^\nu  
B^a_\nu (0) | 0 \rangle = 4S+10R   \label {CT2} 
\end{equation}
and
\begin{equation}
\frac{1}{2} \langle 0 | \partial^\nu B_a^\mu (0) \partial_\mu  
B^a_\nu (0) | 0 \rangle = 4S+10R        \label {CT3}
\end{equation}
respectively. According to the Lorentz gauge 
condition (\ref {LC}), (\ref{CT2}) means
\begin{equation}
R=-\frac{2}{5}S. \label {RS}
\end{equation}

Furthermore, using the definition of the gluon field strength and by
only 
retaining  the contribution of the vacuum intermediate
state \cite{Shakin}, one can easily find that
\begin{eqnarray}
& & \langle 0 | G^a_{\rho \mu}(0) G^b_{\sigma \nu}(0) |0 \rangle \nonumber \\
& \approx & 
\left [ \langle 0 | \partial_\rho B^a_\mu (0) \partial_\sigma  
B^b_\nu (0) | 0 \rangle 
+ \langle 0 | \partial_\mu B^a_\rho (0) \partial_\nu  
B^b_\sigma (0) | 0 \rangle \right ] \nonumber \\ 
& &- \left [  \langle 0 | \partial_\mu B^a_\rho (0) \partial_\sigma  
B^b_\nu (0) | 0 \rangle 
+ \langle 0 | \partial_\rho B^a_\mu (0) \partial_\nu 
B^b_\sigma (0) | 0 \rangle \right ] \nonumber \\ 
& & +\frac{\pi \alpha_s N_c}{3(N_c^2-1)^2} \delta_{ab} 
\left [g_{\rho \sigma} g_{\mu \nu}- g_{\mu \sigma } g_{\rho \nu}\right ] \langle
0| B^2 |0 \rangle ^2  \label {GG1}
\end {eqnarray}
which results in
\begin{equation}
S \approx \frac{5 \langle 0|G^2 |0 \rangle}{288}
- \frac {5N_c \pi \alpha_s}{72(N_c^2-1)}
\langle 0 |B^2 | 0 \rangle ^2.   \label {S1} 
\end{equation}
In addition, combining (\ref{CT3}) with (\ref{CT1}) and (\ref{RS}), 
one can get

\begin{equation}
S=\frac{5
\langle 0| (\partial _\nu B_\sigma^a -
\partial _\sigma B_\nu^a)^2 | 0 \rangle
}{288}
\label {S2} 
\end{equation}
Comparing (\ref{S1}) and (\ref{S2}) leads to

\begin{equation}
\alpha_s  \langle 0 |B^2 | 0 \rangle ^2 \approx 0   \label {GB} 
\end{equation}
provided that  the approximation of the vacuum dominance
in intermediate states is accepted, and that the
equality of
$ \langle 0 |(\partial _\nu B_\sigma^a -
\partial _\sigma B_\nu^a)^2 | 0 \rangle $
and the Abelian part of
$ \langle 0 |G^2 | 0 \rangle $  has been used \cite{Lavelle4}.
The dimension-two condensate
$ \langle 0 |B^2 | 0 \rangle  $ is not gauge invariant.
According to our  estimate of its value,
this term (maybe due to a spontaneous gauge symmetry breaking)
is very small and can be omitted
as compared with the gauge-invariant
gluon condensate $ \langle 0|G^2 |0 \rangle$.

Therefore, by considering (\ref{RS}), (\ref{BB6}) can
also be rewritten as

\begin{eqnarray}
\langle 0 | B^a_\mu (x) B^b_\nu (y) | 0 \rangle &=&
- \frac{\delta_{ab}}{(N_c^2-1)}S
\left [ (x-y)^2 g_{\mu \nu} -\frac{2}{5} (x-y)_\mu 
(x-y)_\nu \right ] + \cdots  \label{BB9}
\end{eqnarray}
with manifest translational invariance. This result was  used
before by Bagan et al.\cite{Bagan} without looking at  
the condensate  $ \langle 0 |B^2 | 0 \rangle  $. 

Now let us focus our attention on discussing the vacuum energy  density
difference and the quark scattering amplitude, which are
related to the nonperturbative gluon propagator.

{\centerline {\bf A. $\xi$-dependence of the vacuum energy density difference}}

The vacuum energy density difference
(the effective
potential of  Coleman and Weinberg \cite{Coleman}) 
 is defined as
the perturbative contribution to the difference
between the energy densities of the physical and
bare vacua.
The condensate contribution to the vacuum energy
difference \cite{Lavelle1} is of interest
because it can tell us something about the energy
dependence on the condensates of the QCD vacuum.
In order to calculate QCD condensate contributions to the
the vacuum energy density difference, we need
quark condensate contributions to the quark self-energy
and the gluon vacuum polarization.
They can be expressed as \cite{Lavelle1}

\begin{equation}
\Sigma ^{<\bar{q} q>}= \sum  \limits_{f}
\frac{(N_c^2 -1) \pi \alpha_s \langle 0|\bar{q}_f q_f |0 \rangle }
{2N_c^2 q^2}
\left [3+\xi - \xi m_f \frac{\qslash}{q^2} \right ]
\end{equation}
and 

\begin{equation}
\Pi _{\mu \nu}^{<\bar{q}q>}= \sum  \limits_{f}
\frac{4 \pi \alpha_s m_f
\langle 0|\bar{q}_f q_f |0 \rangle }{N_c q^2}
 g_{\mu\nu}^{\perp}(q)
\end{equation}
respectively, with $g_{\mu\nu}^{\perp}(q)= g_{\mu\nu}-
q_\mu q_\nu/q^2$.
The form of the gluon condensate contribution depends
on the  choice of the gauge for the external QCD gluon fields.
In the fixed-point gauge,
the contribution to the quark self-energy is\cite{Elias}

\begin{equation}
\Sigma ^{<G^2>}=
\frac{\pi \alpha_s
\langle 0|G^2|0 \rangle  m_f (q^2- m_f \qslash )}
{N_c (q^2 - m_f^2)^3 }
\end{equation}
And, by employing Eq.~(\ref{BB1}), the gluon condensate
contribution to the gluon polarization can be obtained as

\begin{equation}
\Pi _{\mu \nu}^{<G^2>}= \frac{N_c \pi \alpha_s
\langle 0|G^2 |0 \rangle}{12 (N_c^2-1) q^2} \left [ (36+\xi)
g_{\mu\nu}^{\perp}(q)-(12+\frac{18}{\xi})
\frac{q_\mu q_\nu}{q^2} \right].
\label{Puv}
\end{equation}

By using the above self-energies, we obtain the
condensate contributions to the vacuum energy density
in the fixed-point gauge as

\begin{eqnarray}
V(<\bar{q}q>, <G^2>)
&=& \left [
\frac{ 9 \pi \alpha_s N_c
\langle 0|G^2 |0 \rangle}{48}(10-\xi)
+ \sum  \limits_{f}
\frac{6 \pi \alpha_s (N_c^2-1)  m_f
\langle 0|\bar{q}_f q_f |0 \rangle}{N_c}
\right ]  \nonumber \\
& \times & \int \limits_{q^2 < -\mu^2}
(-i) \frac{d^4 q}{(2\pi)^4}
\frac{1}{(q^2+ i \epsilon)^2}, \label{DST}
\end{eqnarray}
which is $\xi$-dependent. The
renormalization point $\mu^2$ is used in QCD to divide
the momentum range into a perturbative
and a nonperturbative regions.
As pointed out by Jackiw \cite{Jackiw}, any
gauge dependence of the effective potential for
a particular operator makes a physical
interpretation questionable. This problem can be avoided 
in covariant gauges. Actually, Lavelle and
Schaden \cite{Lavelle1} got a
gauge  invariant  vacuum energy density difference
by taking into account the ghost condensate contribution
in covariant gauges. With our estimate of
$ \langle 0 |B^2 | 0 \rangle  $ given  in Eq.~(\ref{GB}),
the vacuum energy density difference obtained
in \cite{Lavelle1}
is gauge invariant even if  there exists a  spontaneous
gauge symmetry breaking.

{\centerline {\bf B. Unitarity of the quark scattering amplitude }}

It is well known that the  $\xi$-dependence of
the perturbative gluon propagator
does not carry through to the quark  scattering amplitude.
However, it is quite another story to prove this
for the nonperturbative
gluon propagator. 
In the fixed-point gauge, 
the QCD  condensate contribution to
leading order in $\alpha_s$ to the gluon propagator
can easily be obtained as

\begin{equation}
iD_{\mu\nu}(q)=i \left \{ -\frac{1}{q^2} A_{T}
g_{\mu\nu}^{\perp}(q)+A_{L} \frac{q_\mu q_\nu}{q^4} \right \}
\label{Duv} 
\end{equation}
where

\begin{equation}
A_{T}=\frac{N_c\pi \alpha_s  \langle 0|G^2 |0 \rangle}
{12(N_c^2-1)q^4}
(36+ \xi) +
\sum  \limits_{f}
\frac{4\pi \alpha_s  m_f \langle 0|\bar{q}_f q_f |0 \rangle}{N_c q^2
(q^2-m_f^2)},
\end{equation}

\begin{equation}
A_{L}= \frac{N_c\pi \alpha_s \langle 0|G^2 |0 \rangle}
{12(N_c^2-1) q^4 }
(18\xi +12 \xi^2). 
\end{equation}
The gluon vacuum  polarization used in deriving Eq.~(\ref{Duv})
is given in  Eq.~(\ref{Puv}).
Notice  that the longitudinal term in  (\ref {Puv}) is
both nonzero and $\xi$-dependent.
Hence, the unitarity in
the non-Abelian coupling amplitude cannot be satisfied, 
which stimulates us to try to use a  covariant
gauge for the external
QCD vacuum gluon fields.
At least, by introducing the covariant gauge for
the gluon VEV, the difficulty of two potentially conflicting
gauge conditions for the gluon VEV and the perturbative gluon
fields in QCD can be avoided.

In covariant gauges,
by using the non-local two-gluon VEV given in (\ref {BB9}),
the lowest-dimension gluon condensate contribution to
the gluon propagator can be written as

\begin{equation}
iG_{\mu\nu}^{ <G^2> }(q)=i \left \{
\frac{(25-3\xi)\pi \alpha_s
\langle 0|G^2 |0 \rangle }{24 q^6}g_{\mu\nu}^{\perp}(q)-
\frac{3 \xi^2 \pi \alpha_s 
\langle 0|G^2 |0 \rangle }{8 q^8} q_\mu q_\nu \right \}
\label{Guv1} 
\end{equation}
There is also a longitudinal term in the  gluon polarization
of (\ref {Guv1}), i.e., the Slavnov-Taylor
identities (STI) \cite{STI} is not fulfilled here.
This problem can be fixed by allowing  for 
the presence of the ghost
condensate. In Ref. \cite{Lavelle3}, Lavelle and Schaden got
a transverse gluon vacuum  polarization by taking into account
mixing with equation of motion condensates. Their result
for the gluon vacuum polarization in covariant
gauges with the corrections
of gluon and ghost  condensates as depicted in
Fig.~1(a-c) is

\begin{equation}
\Pi _{\mu\nu}(q^2)=-\frac{N_c \pi \alpha_s (68+3\xi)}{18 (N_c^2-1)q^2}
g_{\mu\nu}^{\perp}
\langle 0| (\partial _\nu B_\sigma^a -
\partial _\sigma B_\nu^a)^2 | 0 \rangle,
\label{Puv2}
\end{equation}
where condensate terms  which vanish due to the equation of motion
are not shown.
It is noteworthy that leading mixed condensate
contributions to the gluon polarization (\ref{Puv2})
have  also been taken into account (the diagrams for the
mixed condensate contributions  are not shown in Fig.~1,
see Ref. \cite{Lavelle3}).
Although the vacuum  polarization (\ref{Puv2}) is transverse, it is
also explicitly $\xi$-dependent. To obtain
a gauge invariant effective gluon propagator,
Lavelle
\cite{Lavelle4} investigated the effects of the
$ \langle 0 | G^2 | 0 \rangle $  condensate on the
effective gluon propagator in quark interactions
by using the pinch technique (PT) in the context of QCD
\cite{Cornwall,Papa}( see  diagrams in
Ref. \cite{Lavelle4} for the PT ).
The effective gluon propagator was finally obtained as

\begin{eqnarray}
iG_{\mu\nu}^{T}(q) &=& i \left  \{ -\frac{1}{q^2}
+ \left [ \frac{34 N_c \pi \alpha_s 
\langle 0|G^2 |0 \rangle }{9 (N_c^2-1) q^6}
\right. \right. \nonumber \\
&-&
\left. \left. \frac{4 \pi \alpha_s}{N_c q^4} \sum  \limits_{f}
m_f \langle 0|\bar{q}_fq_f |0 \rangle
\left ( \frac{1}{q^2-m_f^2}+\frac{1}{2} \frac{m_f^2}{(q^2-m_f^2)^2}
\right ) \right ]
\right \} g_{\mu\nu}^{\perp}(q)
- \xi \frac{q_\mu q_\nu}{q^4}.
\label{Guv4} 
\end{eqnarray}
Here, following Ref. \cite{Lavelle4}, we have identified
$ \langle 0| (\partial _\nu B_\sigma^a -
\partial _\sigma B_\nu^a)^2 | 0 \rangle $
with the Abelian part of
$\langle 0|G^2 |0 \rangle$. The quark condensate
contribution term differs
from that in Eq.~(\ref{Duv})  due to the fact that the 
next-to-leading-order term in the full coefficient
of the $\langle \bar{q}q \rangle$ component of the
nonperturbative two-quark VEV \cite{Elias} is retained.

After the above detailed  discussion on the nonperturbative gluon
propagator, we can now stress our main reason for deriving the
quark interaction potential from the quark scattering amplitude
in covariant gauges for the  external QCD vacuum gluon fields.
First, one can obtain a gauge invariant vacuum energy
density difference in covariant gauges.
However, the $\xi$-dependence of Eq.~(\ref{Puv}) in the
fixed-point gauge 
brings about the explicit $\xi$-dependence in the
vacuum energy density difference as shown in
Eq.~(\ref{DST}),
which raises doubts about its
physical validity.
Second, the transverse portion of the
nonperturbative gluon propagators
in covariant gauges is $\xi$-independent.
In contrast to this, the explicit $\xi$-dependence in
the transverse portion of the nonperturbative gluon propagator
in the fixed-point gauge  results in a quark
scattering amplitude with doubtful physical meaning. All of this
motivate us to employ  the nonperturbative
gluon propagator in covariant gauges to derive the
effective quark interaction potential.

\vspace{0.5cm}

{\centerline {\bf III. QCD CONDENSATE CONTRIBUTIONS TO THE QUARK}} 
{\centerline {\bf  INTERACTION POTENTIALS}}

\medskip

\vspace{0.5cm}

To derive the quark-quark interaction potential from the
nonperturbative gluon propagator, we  write down a proper
scattering amplitude between two quarks as

\begin{equation}
M=(-ig)^2 \bar{\psi}(p_1) \gamma^\mu \frac{\lambda^a}{2}
\psi(p_1^{\prime}) G_{\mu \nu}^{T}(q) \bar{\psi}(p_2) \gamma^\nu
\frac{\lambda^a}{2} \psi (p_2^{\prime})
\end{equation}
with $q=p_1-p_1^\prime=p_2^\prime-p_2$ and the spinors $\psi(p_i)$
being the solution of free quarks. We then obtain the effective
potential in momentum space by carrying out the Breit-Fermi
expansion with the approximation $q_0=0$.
Applying the
three-dimensional Fourier transformation to convert the potential
in momentum space to coordinate space, we finally obtain the total 
effective potential between quarks as

\begin{eqnarray}
U_{qq}(x)=U_{qq}^{\rm{OGEP}}(x)+U_{qq}^{\rm{NP}}(x) \label{U1}
\end{eqnarray}
where $U_{qq}^{\rm{OGEP}}(x)$ is the perturbative
one-gluon-exchange quark potential,

\begin{eqnarray}
U^{\rm{OGEP}}_{qq}(x)
&=&\delta (t) \frac{\lambda_1^a \lambda_2^a}{4} \alpha_s
\left \{ \frac{1}{|\vec{x}|}
-\frac{\pi}{m_1 m_2}\left (\frac{(m_1+ m_2)^2}{2m_1 m_2}  
+\frac{2}{3} \vec{\sigma}_1 \cdot \vec{\sigma}_2\right )\delta (\vec{x})\right.  \nonumber \\
& &+\frac{|\vec{p}|^2}{m_1 m_2|\vec{x}|} -\frac{1}{4m_1m_2|\vec{x}|^3}\left [ \frac{3}{|\vec{x}|^2} 
(\vec{\sigma}_1 \cdot \vec{x})(\vec{\sigma}_2 \cdot \vec{x}) -
(\vec{\sigma}_1 \cdot \vec{\sigma}_2 )\right ]\nonumber \\
& &\left. -\frac{1}{4m_1 m_2 |\vec{x}|^3} 
\left [ (2+\frac{m_2}{m_1})\vec{\sigma}_1
+(2+\frac{m_1}{m_2})\vec{\sigma}_2\right ] 
\cdot (\vec{x} \times \vec{p})\right \},     \label{UOGEP}
\end{eqnarray}                                         
and $U_{qq}^{\rm{NP}}(x)$, the nonperturbative correction
to the perturbative
quark-quark interaction due to the quark, gluon and ghost
condensates, can be expressed as

\begin{eqnarray}
U_{qq}^{\rm{NP}}(x)&=& \delta (t) \frac{\lambda_1^a \lambda_2^a}{4}
\pi \alpha_s^2
\left [A_{3} |\vec{x}|^3+
\left (A_{1}+2C_{1}\right ) |\vec{x}| + 2 C_{-1} |\vec{x}|^{-1}
\right. \nonumber \\
& &\left. +2\sum\limits_f
\left (\tilde{C}_0^{(f)}
+\tilde{C}_{-1}^{(f)}|\vec{x}|^{-1}
\right ){\rm{e}}^{-m_f |\vec{x}|}
\right ].\label{U1NP}
\end{eqnarray}
where

\begin{eqnarray}
A_3=\frac{17N_c  \langle 0|G^2 |0 \rangle  }{108(N_c^2-1)}\left (1+\frac{|\vec{p}|^2}{m_1 m_2}\right ),
\end{eqnarray}

\begin{eqnarray}
A_1&=&\frac{17 N_c  \langle 0|G^2 |0 \rangle  }{72(N_c^2-1)}
\left ( \frac{1}{m_1} +\frac{1}{m_2}\right )^2+
\frac{17 N_c \langle 0|G^2 |0 \rangle  }
{432 m_1 m_2 (N_c^2-1)} ( 8 \vec{\sigma}_1 \cdot \vec{\sigma}_2
-S_{12}) \nonumber \\
&+&\frac{17 N_c  \langle 0|G^2 |0 \rangle  }{144 m_1 m_2 (N_c^2-1)} 
\left [\left ( 2+\frac{m_2}{m_1}\right ) \vec{\sigma}_1
+\left ( 2+\frac{m_1}{m_2} \right ) \vec{\sigma}_2\right ] \cdot 
(\vec{x}\times \vec{p}),
\end{eqnarray}

\begin{eqnarray}
C_1=\left (1+\frac{|\vec{p}|^2}{m_1 m_2}\right )\sum  \limits_{f}
\frac{ \langle 0|\bar{q}_fq_f |0 \rangle}{N_c m_f},
\end{eqnarray}

\begin{eqnarray}
C_{-1}&=&\frac{1}{4N_cm_1m_2}
\sum  \limits_{f}
\frac{\langle 0|\bar{q}_fq_f |0 \rangle}{m_f}
\left \{ 
\frac{(m_1+m_2)^2}{m_1 m_2} +\frac{S_{12}}{3}+\frac{4}{3} 
\vec{\sigma}_1 \cdot \vec{\sigma}_2 \right. \nonumber \\
&+&\left. \left [\left ( 2+\frac{m_2}{m_1}\right ) \vec{\sigma}_1
+\left ( 2+\frac{m_1}{m_2} \right ) \vec{\sigma}_2 \right ] \cdot 
(\vec{x}\times \vec{p})
\right \} 
\end{eqnarray}

\begin{eqnarray}
\tilde{C}_{0}^{(f)}&=&\frac{2}{N_c}
\frac{ \langle 0|\bar{q}_fq_f |0 \rangle}{m_f} 
\left [ 
\frac{1}{2m_f} \left ( 1+\frac{|\vec{p}|^2}{m_1 m_2} \right) \right. \nonumber \\
&+& \left. \frac{m_f(m_1+m_2)^2}{16m_1^2 m_2^2}- 
\frac{m_f}{24m_1 m_2} S_{12}
+\frac{m_f}{12 m_1 m_2} 
\vec{\sigma}_1 \cdot \vec{\sigma}_2 
\right ]
\end{eqnarray}
and

\begin{eqnarray}
\tilde{C}_{-1}^{(f)}&=-&\frac{2}{N_c}
\frac{\langle 0|\bar{q}_fq_f |0 \rangle}{m_f}\left \{
\frac{(m_1+m_2)^2}{8m_1^2 m_2^2}+\frac{S_{12}}{6m_1 m_2}+\frac{1}{6m_1m_2}
\vec{\sigma}_1 \cdot \vec{\sigma}_2 \right.\nonumber \\
&-&\left. \frac{3}{24m_1 m_2}  
\left [\left ( 2+\frac{m_2}{m_1}\right ) \vec{\sigma}_1
+\left ( 2+\frac{m_1}{m_2} \right ) \vec{\sigma}_2 \right ] \cdot 
(\vec{x}\times \vec{p}) \right \}
\end{eqnarray}
with
$S_{12}=3 (\vec{\sigma}_1 \cdot \vec{n})
(\vec{\sigma}_2 \cdot \vec{n})-
\vec{\sigma}_1 \cdot \vec{\sigma}_2$ and
$\vec{n}={\vec{x}}/{|\vec{x}|}$.

For heavy quarkonium systems, where
the potential concept is applicable, the quark-antiquark 
interaction can be obtained by taking the color generators for an
antiquark as $-{\lambda}^T$, i.e.,

\begin{eqnarray}
U^{\rm{Direct}}_{q\bar{q}}(x)&=& U_{qq}(x)
|_{\lambda_1^a \lambda_2^a\rightarrow -\lambda_1^a (\lambda_2^a)^{T}}.
\end{eqnarray}

However, in this case, if 
the quark and antiquark have  the same flavor, the 
annihilation mechanism should  also be taken into 
account. The condensate contributions to this mechanism can be 
sketched in Fig.~2 ( as for leading mixed condensate
contributions and
digrams for the PT, see Refs. \cite{Lavelle3,Lavelle4} ). The  total ${q \bar{q}}$-pair annihilation
potential can be obtained by summing up the contributions of all 
diagrams including nonperturbative ones in Fig.~2 and the
corresponding perturbative one,

\begin{eqnarray}
U^{\rm{Ann}(\rm{Total})}_{q\bar{q}}(x)=U^{\rm{Ann}}_{q\bar{q}}(x)
+ U^{\rm{Ann(NP)}}_{q\bar{q}}(x)
\end{eqnarray}
where $ U^{\rm{Ann}}_{q\bar{q}}(x)$,
the perturbative  $\rm{q\bar{q}}$ pair-annihilation  potential 
in coordinate representation, is

\begin{eqnarray}
U^{\rm{Ann}}_{q\bar{q}}(x)&=& \delta (t) \frac{\alpha_s}{4} \frac{\pi}{16N_cm^2}
(\lambda_1-\lambda_2^{\rm{T}})^2(1- \vec{\tau}_1 \cdot \vec{\tau}_2)
\nonumber \\
& & \times\left \{ (\vec{\sigma}_1+ \vec{\sigma}_2)^2\left (1-\frac{1}{3m^2} 
\vec{\nabla}^2 \right )\delta (\vec{x})
-\frac{4}{m^2}\left [(\vec{\sigma}_1 \cdot \vec{\nabla})(\vec{\sigma}_2 \cdot \vec{\nabla}) \right. \right.\nonumber \\
& &\left. \left.-\frac{1}{3} \vec{\sigma}_1 \cdot \vec{\sigma}_2 \vec{\nabla}^2 \right ] \delta(\vec{x})\right \}.
\end{eqnarray}
and

\begin{eqnarray}
U^{\rm{Ann(NP)}}_{q\bar{q}}(x) &=& \frac{\pi \alpha_s}{m^2}
\left \{ \frac{N_c}{(N_c^2-1)}
\frac{17 \langle 0|G^2 |0 \rangle  }{72 m^2}
\right.  \nonumber \\
& & \left.
+\frac{1}{N_c}
\sum  \limits_{f}
\frac{m_f \langle 0|\bar{q}_fq_f |0 \rangle}{(4m^2-m_f^2)^2}
(8m^2-m_f^2) \right \}
U^{\rm{Ann}}_{q\bar{q}}(x).
\end{eqnarray}

\vspace{0.5cm}

{\centerline {\bf IV. SUMMARY AND DISCUSSION }}

\medskip

\vspace{0.5cm}

In this paper, we discussed the condensate
contributions to the
gluon propagator which is then used to derive the nonperturbative
contributions to the quark potentials.
To help the reader to understand
what is new in this paper, we
summarize some important points:

(1) We estimated
the value of $ \langle 0 |B^2 | 0 \rangle ^2 $
in the approximation of the vacuum dominance in intermediate states
and found  that this non-gauge-invariant condensate can be
omitted as  compared with the gauge-invariant gluon
condensate contribution, which is of crucial importance for 
having  gauge invariant vacuum energy density differences 
\cite{Lavelle1}.

(2) We gave a detail discussion about the
nonperturbative gluon propagator and showed 
that it is essential  to adopt 
covariant gauges in order to obtain a
gauge invariant vacuum energy density difference
and to retain the unitarity of the quark scattering
amplitude.

(3) The gauge-invariant nonperturbative
contributions to the one-gluon
exchange quark potentials were  obtained by employing
the nonperturbative gluon propagator, with
the validity of the unitarity of the quark scattering amplitude
from which the quark potential is derived.

In the fixed-point gauge, some uncertainties in
the nonperturbative  calculation such as 
three-point fermionic Green function $G^{\mu}_{\alpha\beta}
(p^\prime, p, q)$ are unavoidable \cite{YSL}. To overcome this,
one can complete the calculation in  covariant gauges \cite{YSLH}.
Of course, it is not enough to judge whether the result
in any gauge for the external QCD vacuum gluon fields is better
than the one in others only by calculating
gauge-dependent objects
such as quark-gluon n-point Green functions.
However, we should
accept the fact that the explicit $\xi$-dependence of
the energy density difference in the fixed-point gauge
is incompatible with the physical meaning of
a gauge invariant object. In the context of the OPE, the
fixed-point gauge is  very convenient. But, as we have seen here,
one should be very careful in using this gauge.

There are many possible applications of the obtained quark potentials,
for instance, in the study of the nonperturbative 
effect in the spectra
of $J/\Psi$ and ${\Upsilon}$, 
especially to improve the spin splitting for
these systems. In addition, this work is
intended to serve as a step forward in the direction of solving
long-standing problems in light baryon spectroscopy such as the
energy level order between the positive- and negative- parity
partner states, in particular, Roper resonance puzzle, and
the baryon spin-orbit structure puzzle.

As an extension of this work, we will verify whether
the potentials obtained here can be used to improve the
hadronic  spectra and hadronic properties of
$J/{\Psi}$ and ${\Upsilon}$ families by including perturbative
closed-loop contributions in the same order of $\alpha_s$ as shown
by Gupta et al. \cite{Gupta},
Fulcher \cite{Fulcher} and Pantaleone et al. \cite{Pant}.
Moreover, physically relevant results, such
as the effective quark-quark interaction potential,  should
be gauge-independent, i.e., the result in the fixed-point
gauge  should be the same as  that in  covariant gauges.
The discrepancy  between the present
result and that of Ref. \cite{SPN} deserves further study.
(1) We plan to consider all contributions from
the gauge invariant set of diagrams such as gluon
propagator, ghost propagator and quark-gluon vertex
corrections;
(2) We will verify
whether the gauge-independent quark potential
can be obtained  by using the gauge invariant definition of 
the quark potential from the Wilson loop.

\begin{center}
{\bf
ACKNOWLEDGEMENTS}
\end{center}
We are very grateful to Dr. M. Lavelle for
sending us valuable comments on a preliminary
version of this paper. This work was supported in part
by Fondecyt (Chile) postdoctoral fellowship 3990048, 
by Fondecyt (Chile) grant 1990806 and by a C\'atedra 
Presidencial (Chile), and also by  Natural Science
Foundation of China grant 19875024.

\newpage

\vspace{0.5cm}
\begin{figure}[htb]
\begin{center}
\leavevmode {\epsfysize=10cm \epsffile{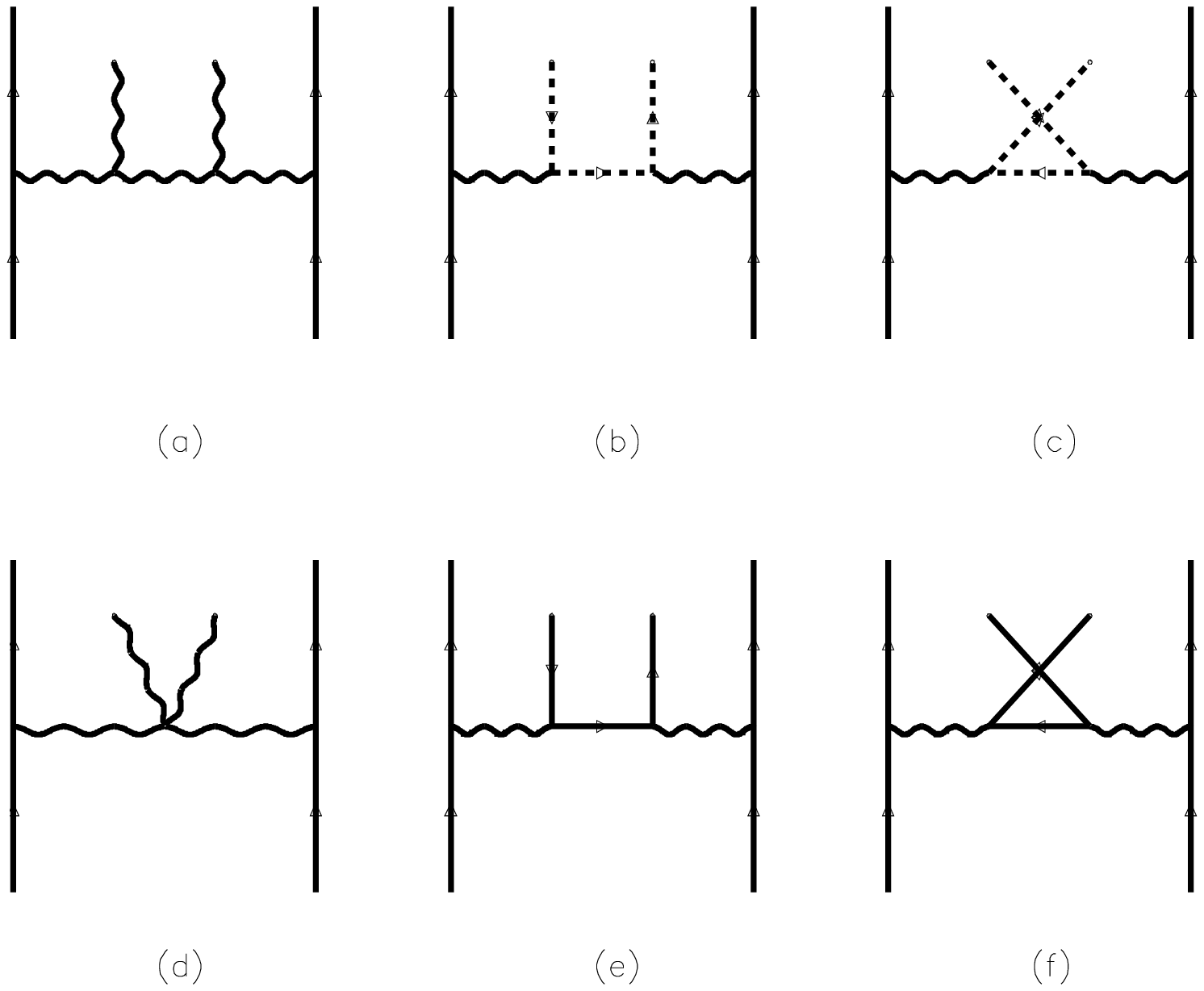}} 
\end{center}
\caption[*]{\baselineskip 13pt 
The Feynman diagrams for the  contributions of  the nonperturbative  
corrections to perturbative quark-quark potential  in the 
one-gluon exchange 
approximation with  the lowest dimensional gluon, ghost and quark
condensates.}
\label{sy1}
\end{figure}

\vspace{0.5cm}
\begin{figure}[htb]
\begin{center}
\leavevmode {\epsfysize=10cm \epsffile{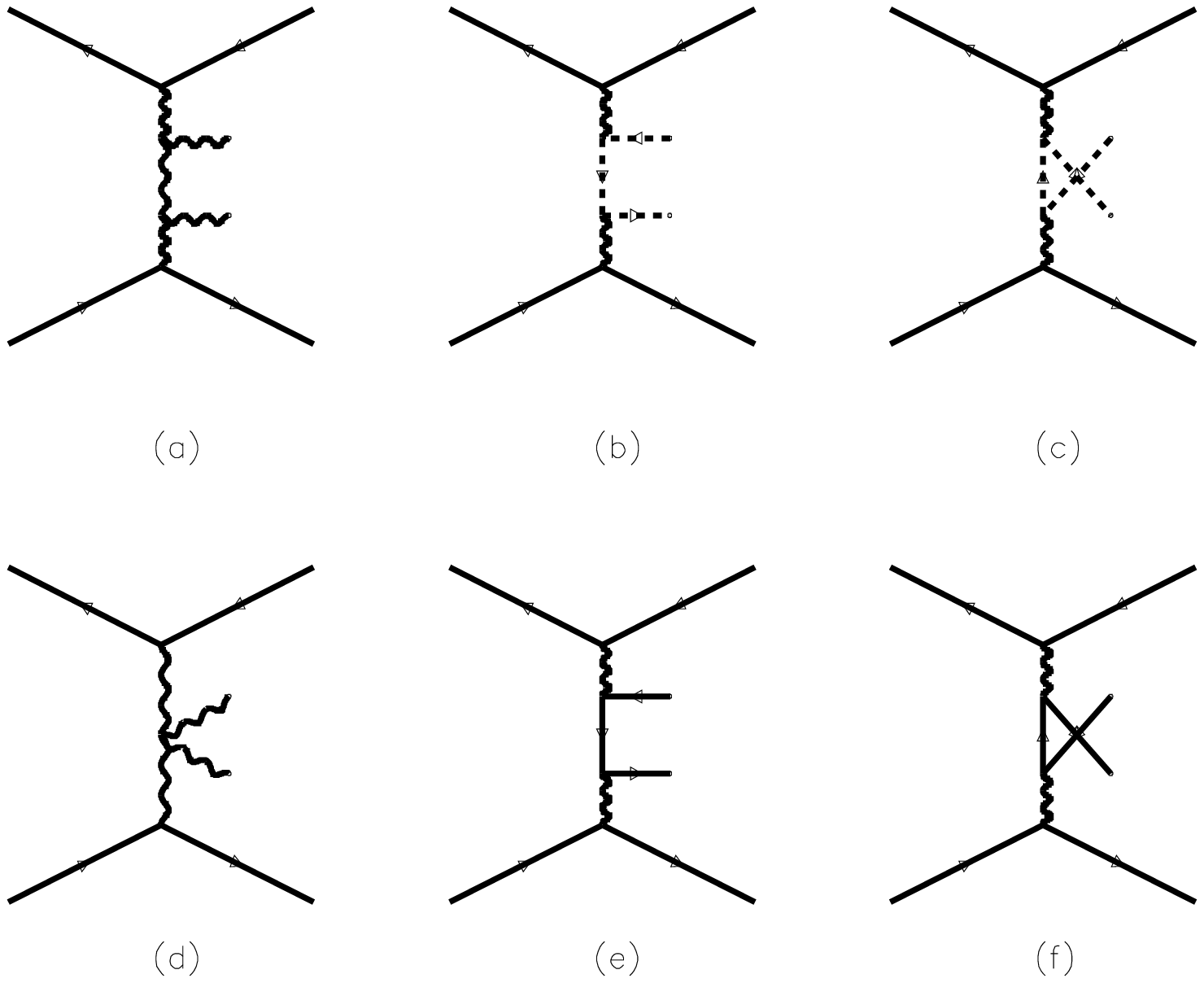}} 
\end{center}
\caption[*]{\baselineskip 13pt 
The Feynman diagrams for the contributions of  the nonperturbative  
corrections to perturbative $q \bar{q}$-pair annihilation potential  
in the  one-gluon exchange approximation with  the lowest 
dimensional gluon, ghost and quark condensates.}
\label{sy2}
\end{figure}

\end{document}